\documentclass[aps,prb,twocolumn,groupedaddress,showpacs,amsmath,amssymb,amsfonts,floatfix]{revtex4}

\usepackage{graphicx}

\begin{document}
\title{Magnetization Plateau of Classical Ising Model on Shastry-Sutherland Lattice}

\author{Ming-Che Chang}
\affiliation{Department of Physics, National Taiwan Normal
University, Taipei, Taiwan}

\author{Min-Fong Yang}
\email{mfyang@thu.edu.tw}
\affiliation{Department of Physics,
Tunghai University, Taichung, Taiwan}

\date{\today}

\begin{abstract}
We study the magnetization for the classical antiferromagnetic
Ising model on the Shastry-Sutherland lattice using the tensor
renormalization group approach. With this method, one can probe
large spin systems with little finite-size effect. For a range of
temperature and coupling constant, a single magnetization plateau
at one third of the saturation value is found. We investigate the
dependence of the plateau width on temperature and on the strength
of magnetic frustration. Furthermore, the spin configuration of
the plateau state at zero temperature is determined.
\end{abstract}


\maketitle

\section{Introduction}

\begin{figure}
\includegraphics[width=1.5in]{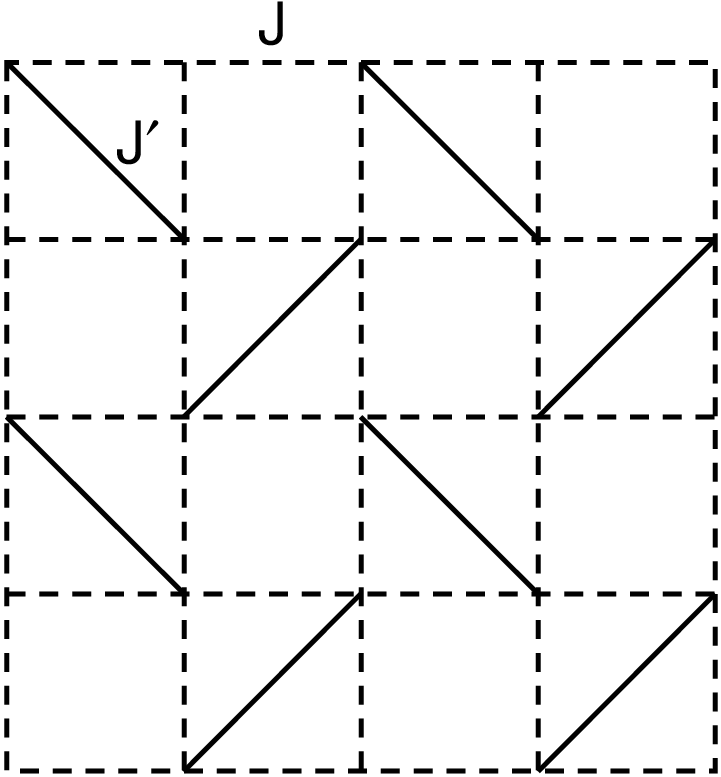}
\caption{The Shastry-Sutherland lattice. $J$ bonds (dashed lines)
are the exchange couplings along the edges of the squares and
$J^\prime$ bonds (solid lines) are the diagonal intra-dimer
couplings.}\label{fig:SSL}
\end{figure}

The frustrated spin systems have attracted much attention over
last decades since very rich physics can appear in these
systems.~\cite{review_frustration} Some interest in such systems
is concentrated on fascinating sequence of magnetization plateaus
at fractional values of the saturation magnetization, which was
first observed in two-dimensional spin-gap material
SrCu$_2$(BO$_3$)$_2$.~\cite{SrCuBO} This compound can be described
well by spin-1/2 antiferromagnetic Heisenberg model on the
frustrated Shastry-Sutherland lattice (or the orthogonal-dimer
lattice),~\cite{Shastry81} as shown in Fig.~\ref{fig:SSL}. Besides
the previously discovered plateaus at $1/3$, $1/4$ and $1/8$ of
the saturated magnetization, evidence in favor of more fractional
magnetization plateaus down to values as small as $1/9$ has been
reported recently.~\cite{Sebastian07,Levy08,Takigawa07} Stimulated
by the discovery of magnetization plateaus, various theoretical
and experimental explorations have been devoted to the properties
of the Shastry-Sutherland model in a magnetic
field.~\cite{Miyahara03,Dorier08,Schmidt08,Abendschein08}

Similar phenomena of magnetization plateaus is also observed in
rare-earth tetraborides RB$_4$. The magnetic ions of these
compounds are again located on a lattice that is topologically
equivalent to the Shastry-Sutherland
lattice.~\cite{Yoshii07,Yoshii08,Michimura06,Yoshii06,Iga07,Gabani08,Siemensmeyer08}
In particular, magnetization plateaus at small fractional values
($1/7, 1/9 \dots$ of the saturation magnetization) are reported in
the compound TmB$_4$.~\cite{Gabani08,Siemensmeyer08} Because fully
polarized state can be reached for experimentally accessible
magnetic fields, this compound allows exploration of its complete
magnetization process. Note that, due to large total magnetic
moments of the magnetic ions, this compound can be considered as a
classical system. Moreover, because of strong crystal field
effects, the effective spin model for TmB$_4$ has been suggested
to be described by the spin-1/2 Shastry-Sutherland model under
strong Ising (or easy-axis) anisotropy.~\cite{Siemensmeyer08}
Thus, studying the Ising limit is the first step toward a complete
understanding of the magnetization process for this material.

In the presence of a finite magnetic field $h$, the total energy
of the antiferromagnetic Ising model on the Shastry-Sutherland
lattice is given by
\begin{equation}
E(\{s_i\})=J\sum_{< i,j >} s_i s_j + J^\prime\sum_{\ll i,j \gg}
s_i s_j - h \sum_i s_i \; , \label{eq:total_eng}
\end{equation}
with exchange couplings $J$, $J^\prime\geq 0$. Here, $s_i = \pm
1/2$ denotes the $z$-component of a spin-1/2 degree of freedom on
site $i$ of the square lattice. The first sum extends over all
nearest neighbor bonds, and the second sum runs over next-nearest
neighbor bonds in every second square, as indicated in
Fig.~\ref{fig:SSL}. Even for this simplified case, different
conclusions for the magnetization curve have been reached. In
Ref.~\onlinecite{Siemensmeyer08}, a single plateau at $1/2$ of the
saturation magnetization is found based on analyzing a finite
system with 16 spins only. However, when larger system sizes up to
$18\times18$ spins are considered, a distinct plateau at $1/3$ of
the saturation magnetization is obtained.~\cite{Meng08} The
discrepancy may come from the effect of finite lattice sizes. As
noted by the authors of Ref.~\onlinecite{Meng08}, for finite
systems, inappropriate lattice sizes and boundary conditions can
frustrate certain magnetization patterns, and hence lead to rather
different magnetization curves which do not correctly represent
the behavior in the thermodynamic limit. For example, the plateau
at $1/3$ of the saturation magnetization is not allowed for
systems of $4\times4$ and $8\times8$ spins, even though it does
describe the true magnetization process for systems in the
thermodynamic limit.

In order to check theoretically if other reported magnetization
plateaus at small fractional values can be stabilized in the
current model, unbiased large-scale calculations are called for.
This is because the unit cells of magnetization profiles inside
high-commensurability plateaus are usually quite large,
calculations for systems of finite sizes may prevent reliable
predictions for these cases. Therefore, to avoid the frustration
for certain magnetization plateaus coming from geometric
constraints, and in particular to uncover the possibility of
plateaus at small fractional values, analyzing systems of large
enough sizes are necessary.

Lately, based on ideas from quantum information theory, the tensor
renormalization group (TRG) method is developed,~\cite{Levin07}
which can efficiently calculate quantities of classical systems of
very large sizes. This technique can in principle be applied to
any classical lattice with local interactions as long as the
partition function can be expressed as a tensor
network.~\cite{Shi06} Because the accuracy can be systematically
improved by increasing the cutoff on the index range of the
tensors, highly precise quantities can be calculated under the TRG
approach even in the thermodynamic
limit.~\cite{Levin07,Hinczewski08,Gu08} Therefore, the TRG method
is one of the most suitable ways to study the magnetization
process of the classical frustrated spin systems in the
thermodynamical limit.

In the present work, the magnetization process of the spin-1/2
Shastry-Sutherland model in the Ising limit is investigated by
employing the TRG approach.~\cite{Levin07,Hinczewski08,Gu08} We
find that the magnetization curve exhibits exactly one plateau at
$1/3$ of the saturation value. Our results are in accordance with
the findings in Ref.~\onlinecite{Meng08}. Furthermore, phase
diagrams in the ($h, T$) plane for a typical magnetic coupling
ratio $J^\prime/J = 1$ and in the ($h, J^\prime$) plane for a
particular temperature $T/J = 0.2$ are obtained. Since there is no
evidence for the presence of any additional plateaus for the
spin-1/2 Shastry-Sutherland model in the Ising limit, to explain
the experimental results, one must go beyond this simple model.

This paper is organized as follows. In Sec.~II, the TRG approach
is outlined briefly. In Sec.~III, we apply this method to
investigate the magnetization process of the Shastry-Sutherland
model in the Ising limit. The spin configuration of the plateau
state at zero temperature is discussed in Sec.~IV. Sec.~V is the
conclusion.

\section{TRG formulation}

\begin{figure}
\includegraphics[width=1.5in,angle=90]{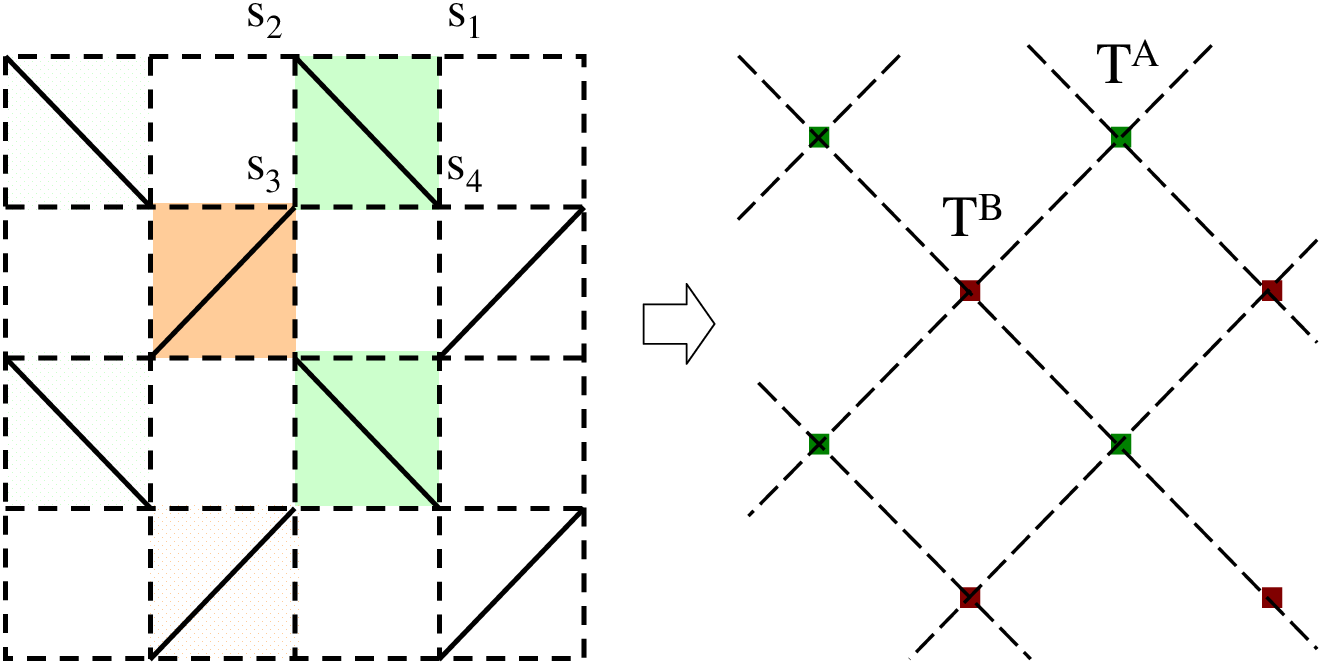}
\caption{(Color online)  Checkerboard decomposition of the
Shastry-Sutherland lattice and the corresponding tensor
network.}\label{fig:TN}
\end{figure}

Before applying the TRG method of Levin and Nave,~\cite{Levin07}
we first explain how to express the partition function of the
present model as a tensor network. One possible way is to rewrite
the total energy in Eq.~(\ref{eq:total_eng}) as a summation over
the energies of plaquettes with diagonal bonds.~\cite{note1} The
energy of, say, the plaquette $A$ with spins $s_1, s_2, s_3, s_4$
on its corners is given by (see Fig.~\ref{fig:TN})
\begin{eqnarray}
\epsilon^A(s_1,s_2,s_3,s_4) &=&
J(s_1s_2 + s_2s_3 + s_3s_4 + s_4s_1)\nonumber\\
&+& J^\prime s_2s_4 - \frac{h}{2}(s_1 + s_2 + s_3 + s_4) \; .
\end{eqnarray}
The rank-four tensors are defined as the Boltzmann weights for
these plaquettes. For example,
\begin{equation}
T^A_{\alpha_1,\alpha_2,\alpha_3,\alpha_4}
=\exp[-\beta\epsilon^A(s_1,s_2,s_3,s_4)]
\end{equation}
with $\beta$ being the inverse temperature and the indices
$\alpha_i \equiv s_i+3/2$ running over 1 and 2. Afterwards, the
partition function can be rewritten as a sum of tensor products in
the following way,
\begin{eqnarray}
Z&=& \sum_{\{s_i\}}e^{-\beta E(\{s_i\})}\nonumber \\
&=& {\rm tTr}\left(T^A T^B\cdots\right) \; .
\end{eqnarray}
Here the tensor trace (tTr) means that all indices on the
connected links in the tensor products are summed over. As a
result, the partition function of the Ising model on the
Shastry-Sutherland lattice is transformed to a tensor network as
shown on the right-hand side of Fig.~\ref{fig:TN}.

As discussed in Refs.~\onlinecite{Levin07,Hinczewski08,Gu08}, the
tensor network can be coarse-grained in an iterative fashion to
reduce the load of computation. At the mean time, the accuracy can
be controlled by a parameter of cutoff $D_{cut}$. Here we outline
the process briefly. Each step of the renormalization consists of
two operations: rewiring and decimation. After one step of the
renormalization, the number of sites in the tensor network is
reduced by half. Eventually, the system is reduced to only four
sites (four $T$'s) and the partition function can be calculated
with ease.

\begin{figure}
\includegraphics[width=3in]{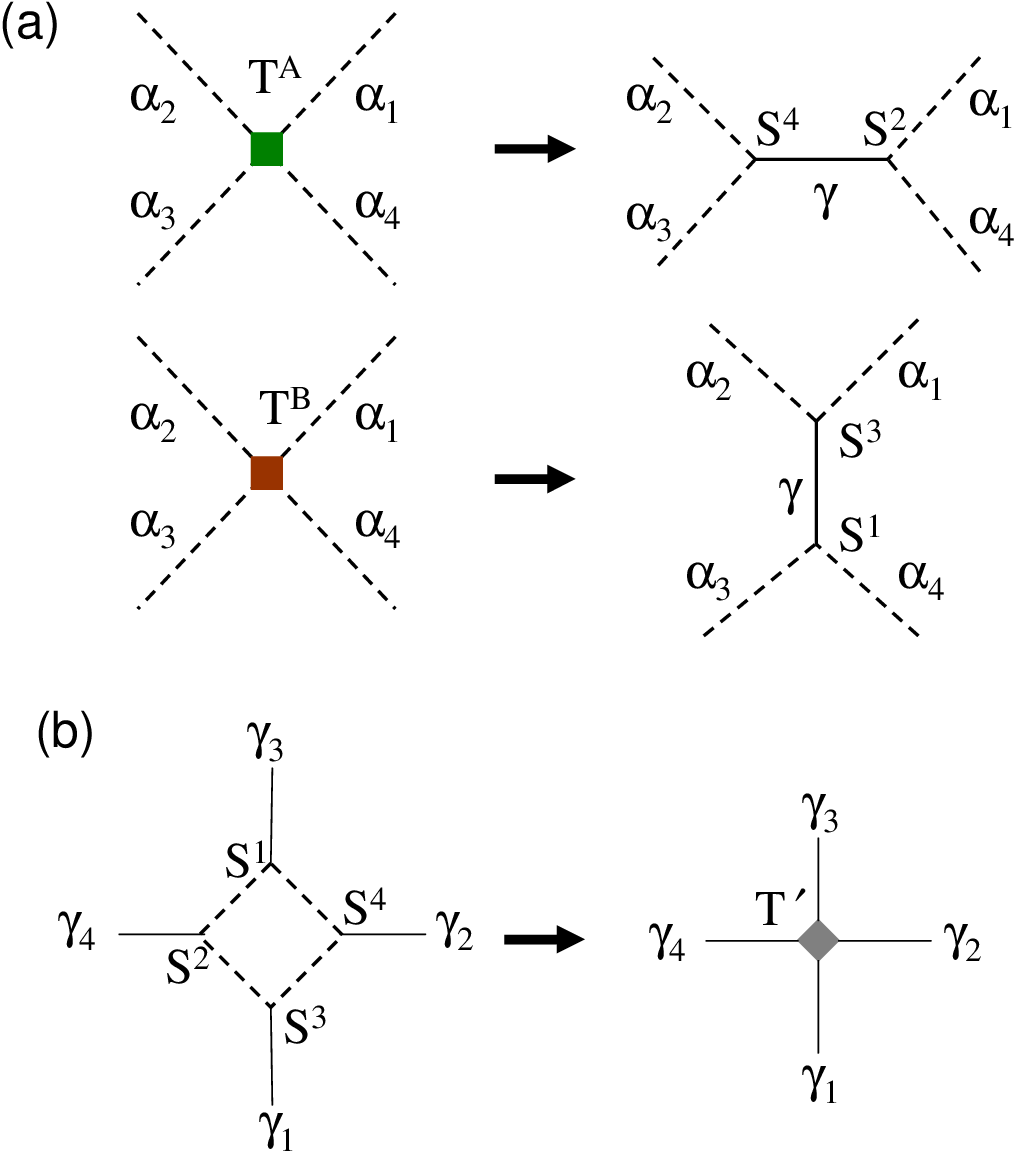}
\caption{(a) Rewiring: the original rank-four tensors are
decomposed to two rank-three tensors. (b) Decimation: the new
tensor $T^\prime$ is obtained by summing over the indices around
the square. }\label{fig:TRG-1}
\end{figure}

{\it Rewiring} -- By viewing the rank-four tensor as a matrix, say
$M_{(\alpha_2,\alpha_3),(\alpha_4,\alpha_1)}
=T^A_{\alpha_1,\alpha_2,\alpha_3,\alpha_4}$, and with the help of
singular value decomposition (SVD), $M=U\Lambda V^\dagger$, the
rank-four tensor can be decomposed to two rank-three tensors. That
is [see Fig.~\ref{fig:TRG-1}(a)],
\begin{eqnarray}
T^A_{\alpha_1,\alpha_2,\alpha_3,\alpha_4} &=&
\sum_{\gamma=1}^{D^2}
S^{4}_{(\alpha_2,\alpha_3),\;\gamma}S^{2}_{(\alpha_4,\alpha_1),\;\gamma} \; , \nonumber \\
T^B_{\alpha_1,\alpha_2,\alpha_3,\alpha_4} &=&
\sum_{\gamma=1}^{D^2}
S^{3}_{(\alpha_1,\alpha_2),\;\gamma}S^{1}_{(\alpha_3,\alpha_4),\;\gamma}
\; . \label{tatb}
\end{eqnarray}
Here
$S^{4}_{(\alpha_2,\alpha_3),\;\gamma}=\sqrt{\lambda_\gamma}U_{(\alpha_2,\alpha_3),\;\gamma}$,
$S^{2}_{(\alpha_4,\alpha_1),\;\gamma}=\sqrt{\lambda_\gamma}V^\dagger_{\gamma,\;(\alpha_4,\alpha_1)}$
(similarly for $S^{3}$ and $S^{1}$), in which $\lambda_\gamma$ are
the singular values, and $U$, $V$ are the unitary matrices in SVD.
If each index of the original rank-four tensor has $D$ possible
values, then there should be $D^2$ terms in the summation of
Eq.~(\ref{tatb}). In practice, the tensor is approximated by
keeping only the largest $D_{cut}$ singular values and the
corresponding singular vectors. Apparently, the cutoff needs to be
chosen such that the result converges with little
$D_{cut}$-dependence.

\begin{figure}
\includegraphics[width=1.5in,angle=90]{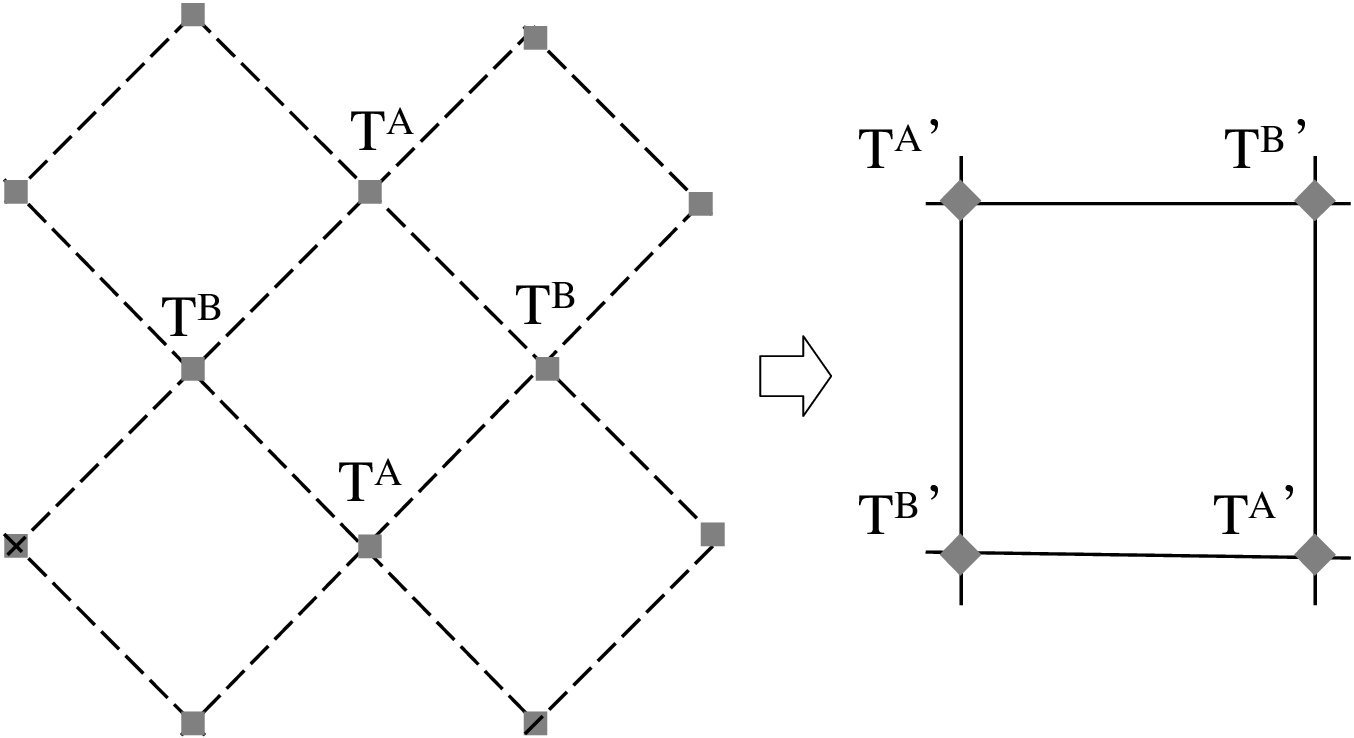}
\caption{Under the TRG procedure, a tensor network is transformed
into a coarse-grained tensor network.}\label{fig:TRG-2}
\end{figure}

{\it Decimation} -- After rewiring, the dashed lines in
Fig.~\ref{fig:TRG-1}(a) can be closed to build a new rank-four
tensor, $T^\prime$ [see Fig.~\ref{fig:TRG-1}(b)]. This is achieved
by the following operation,
\begin{equation}
T^\prime_{\gamma_1,\gamma_2,\gamma_3,\gamma_4}= {\rm Tr}
\left(\textsf{S}^4_{\gamma_2} \textsf{S}^3_{\gamma_1}
\textsf{S}^2_{\gamma_4} \textsf{S}^1_{\gamma_3}\right) \; ,
\end{equation}
where the square matrices
$\left(\textsf{S}^k_\gamma\right)_{\alpha,\alpha^\prime} \equiv
S^k_{(\alpha,\alpha^\prime),\gamma}$. After such a contraction,
one obtains a new tensor network that is half of the size (see
Fig.~\ref{fig:TRG-2}). Afterwards, the renormalization can be
carried out iteratively until there are only four sites left.

We note that, to prevent the computation from diverging, one needs
to normalize the new rank-four tensor at each step of the
renormaliztion. At the beginning, we factor out the maximal value
$W_0$ of the tensor elements of $T^{A/B} \equiv T^{A/B}_0$ to
obtain a normalized tensor $\tilde{T}^{A/B}_0$. After the first
step of the renormalization-group (RG) transformation on
$\tilde{T}^{A/B}_0$, a renormalized tensor $T^\prime \equiv T_1$
is reached. Now we choose the normalization factor to be
$W_1=\lambda^A_{max} \lambda^B_{max}$ such that $T_1 = W_1
\tilde{T}_1$, where $\lambda^A_{max}$ and $\lambda^B_{max}$ are
the largest singular values of the two decompositions in
Eq.~(\ref{tatb}).

The factorization and RG transformation are then iterated, so that
at the $n$th step we have a tensor $T_n = W_n \tilde{T}_n$. Thus,
for the Shastry-Sutherland lattice of $N=2^{n+3}$ sites (and with
$N/2$ tensors in the original tensor network), after $n$ steps of
the RG transformation, one has
\begin{eqnarray}
Z&=& {\rm tTr}\left(T_0^A T_0^B \cdots T_0^B \right)\nonumber\\
&=& W_0^{N/2} W_1^{N/4} \cdots W_{n}^{N/2^{n+1}} {\rm
tTr}\left(\tilde{T}_n^A \tilde{T}_n^B \tilde{T}_n^A \tilde{T}_n^B
\right) \; . \label{Z}
\end{eqnarray}
Since the last tensor-trace term in Eq.~(\ref{Z}) remains finite,
its contribution to the free energy can be neglected for a large
enough system. The free energy per site thus becomes
\begin{equation}
f = -\frac{1}{\beta}\frac{\ln Z}{N} \simeq
-\frac{1}{\beta}\sum_{i=0}^n \frac{1}{2^{i+1}}\ln W_i \; .
\label{f}
\end{equation}
Once the free energy is obtained, one can proceed to calculate the
magnetization. The results are shown and discussed in the
following sections.

\section{Numerical results}\label{numer}

In this section, we present the numerical results on the
magnetization plateau and related phase diagrams. Throughout the
region being explored, we find only one magnetization plateau at
$m/m_s=1/3$, where $m$ denotes the magnetization and $m_s$ its
saturation value. Unless otherwise mentioned, the size of the
system is $2^{10}\times 2^{10}$ with periodic boundary condition.
That is, the number of steps of the RG transformation in
Eqs.~(\ref{Z}) and (\ref{f}) is $n=17$. The temperature $T$ and
the strength of magnetic frustration $J^\prime$ are measured in
units of $J$.

\begin{figure}
\includegraphics[width=3in]{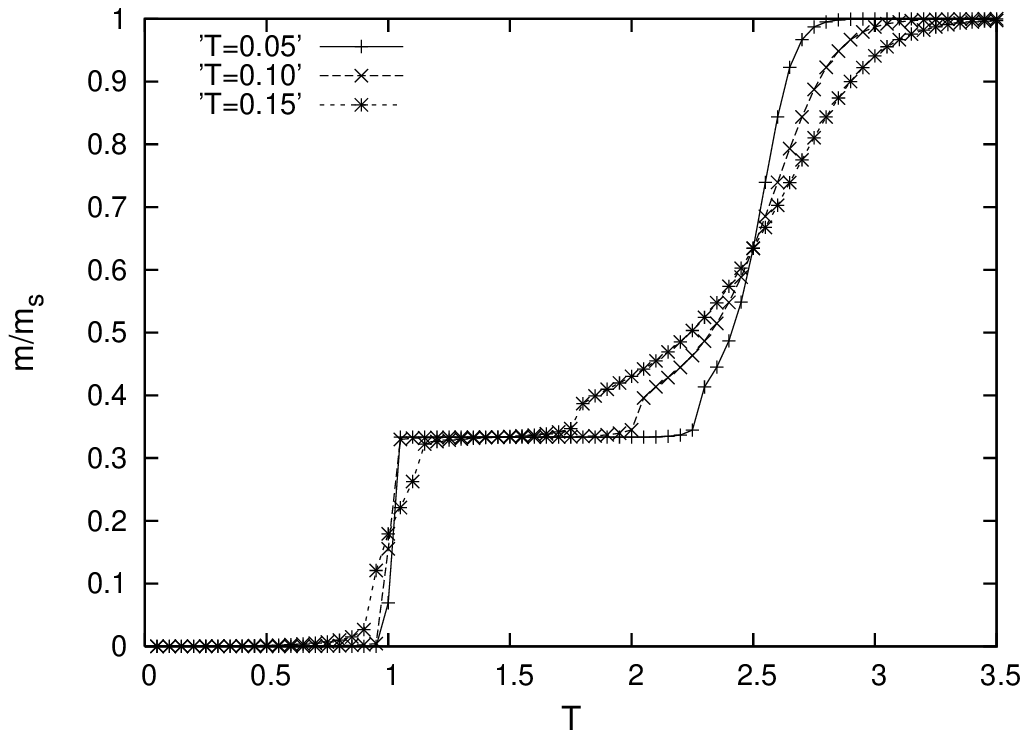}
\caption{Magnetization curves for three different temperatures.
The size of the system is $2^{10}\times 2^{10}$. The parameters
are $J^\prime =1$ and $D_{cut}=18$.}\label{fig:M_curve}
\end{figure}

Fig.~\ref{fig:M_curve} is a typical diagram of the magnetization
curves for $J^\prime =1$. The curves for three temperatures
($T=0.05$, $0.1$ and $0.15$) are shown. The size of the system is
$2^{10}\times 2^{10}$ and the cutoff $D_{cut}=18$. Current result
converges well against further increase of the system size and the
cutoff. For example, for $T=0.05$, a larger system with
$2^{15}\times 2^{15}$ ($D_{cut}=18$) yields a result agrees to the
sixth decimal place for the most part of the curve. A larger
cutoff $D_{cut}=24$ (system size $2^{10}\times 2^{10}$) shows
similar accuracy. The result is slightly less accurate near the
edges of the magnetization plateau but still shows no visible
difference from the $T=0.05$ curve in Fig.~\ref{fig:M_curve}.
Compared to other methods, the TRG method is both accurate and
efficient for very large systems.

\begin{figure}
\includegraphics[width=3in]{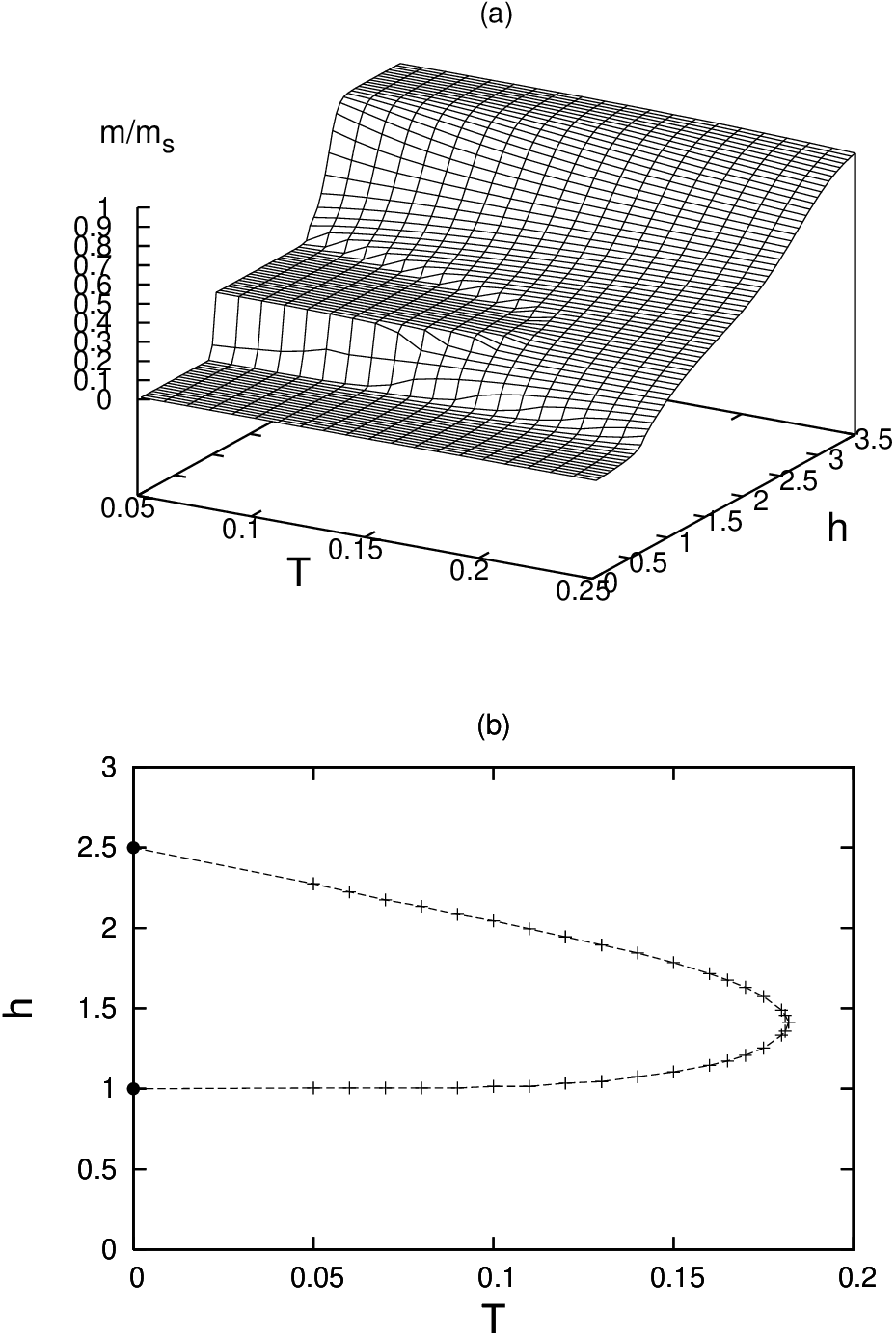}
\caption{(a) Magnetization versus temperature $T$ and magnetic
field $h$. The parameters are $J^\prime =1$ and $D_{cut}=18$. (b)
Phase diagram of the magnetization plateau. The theoretical values
of the critical fields at zero temperature are denoted by filled
circles. }\label{fig:T-h_diagram}
\end{figure}

A more complete scan of the temperature can be found in
Fig.~\ref{fig:T-h_diagram}(a). Over the whole range of
calculation, there is only one plateau at $m/m_s=1/3$. Its width
gradually shrinks to zero near temperature $T=0.18$. The
corresponding phase diagram for the 1/3-plateau is shown in
Fig.~\ref{fig:T-h_diagram}(b). The extent of the plateau is
determined by the locations of maximum slope near its edges, which
will be denoted as $h_{c,1}$ and $h_{c,2}$ for the lower and the
higher critical fields respectively. In
Fig.~\ref{fig:T-h_diagram}(b), we have added the theoretical
critical fields $(h_{c,1},h_{c,2})=(1,5/2)$ at zero temperature
(details later). One can see that the numerical result does
extrapolate to the theoretical values as temperature decreases.

\begin{figure}
\includegraphics[width=3in]{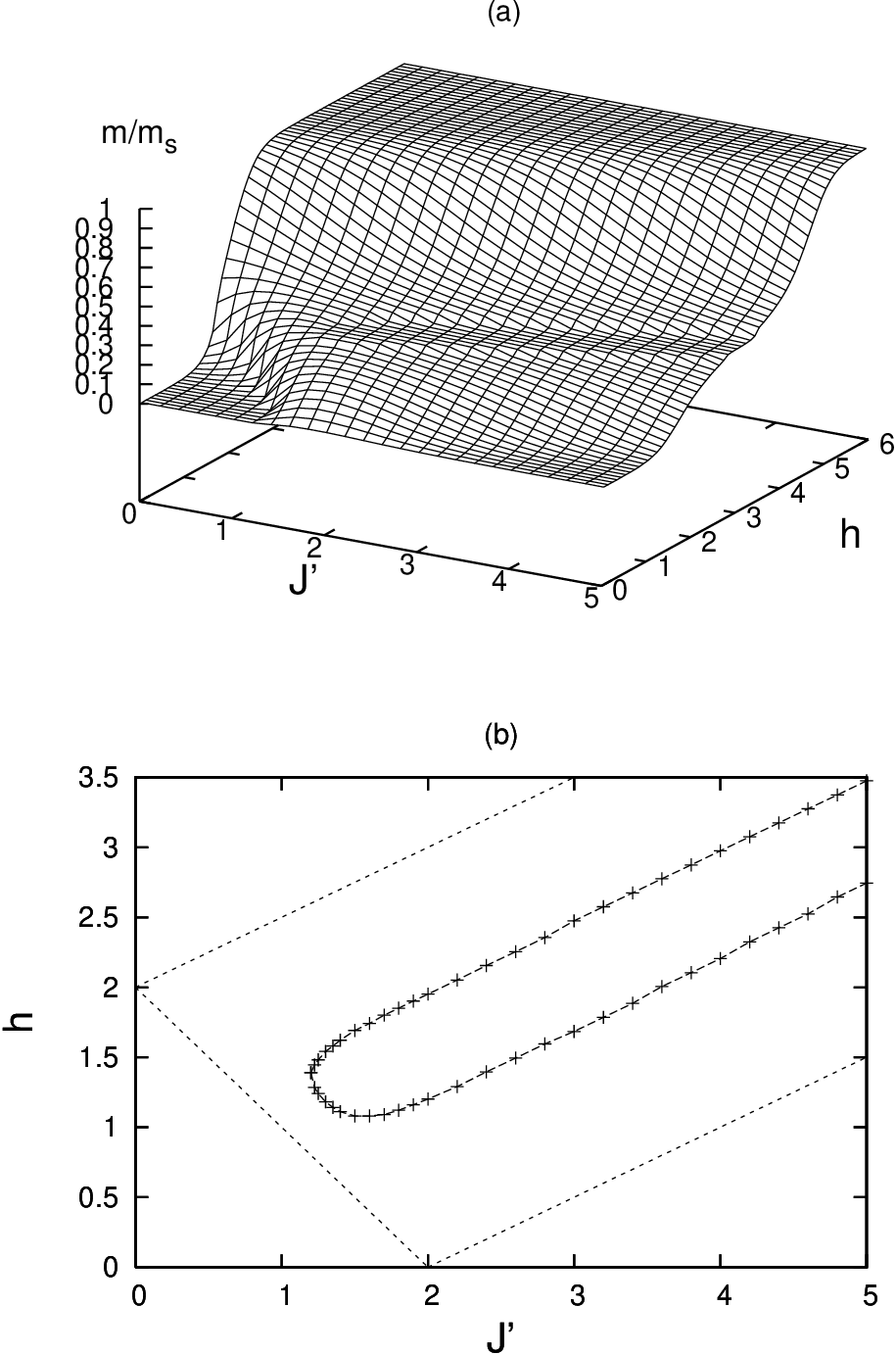}
\caption{(a) Magnetization versus frustration $J^\prime$ and
magnetic field $h$. The parameters are $T=0.2$ and $D_{cut}=18$.
(b) Phase diagram of the magnetization plateau. Dashed lines are
the theoretical phase boundaries at zero
temperature.}\label{fig:J'-h_diagram}
\end{figure}

In Fig.~\ref{fig:J'-h_diagram}(a), we show another scan of the
magnetization with respect to $J^\prime$ and $h$ at $T=0.2$. At
this temperature, there is no plateau for small frustration. The
plateau appears when $J^\prime$ is slightly larger than 1. One can
see that the widths of the plateaus remain roughly the same for
$J^\prime >2$. Their positions appear to shift linearly with
respect to the strength of the frustration $J^\prime$. One can see
this clearly in the phase diagram of
Fig.~\ref{fig:J'-h_diagram}(b). The plateaus are again determined
by the locations of maximum slope. The characters of this phase
diagram at finite temperature are inherited from its counterpart
at zero temperature (details later), which is also plotted in
Fig.~\ref{fig:J'-h_diagram}(b) for comparison. The plateaus at
zero temperature indeed exhibit a constant width at large
frustration and a linear shift of the plateau position. Such a
behavior will be explained in the next section.

\section{Magnetization plateau at zero temperature}

When the temperature is zero, the system is in the ground state.
If the spin configuration of the ground state is known, then the
Ising energy of the system can be calculated analytically.
Afterwards, by comparing the ground state energies at different
parameters, one can determine the phase boundaries in the
parameter space. In this section, we will consider three regimes
of magnetization: the unmagnetized state ($m=0$), the state of the
$1/3$-plateau, and the fully-magnetized state ($m/m_s=1$). It will
be shown that the phase boundaries being determined are consistent
with the numerical results reported in Sec.~\ref{numer}.

\begin{figure}
\includegraphics[width=3in]{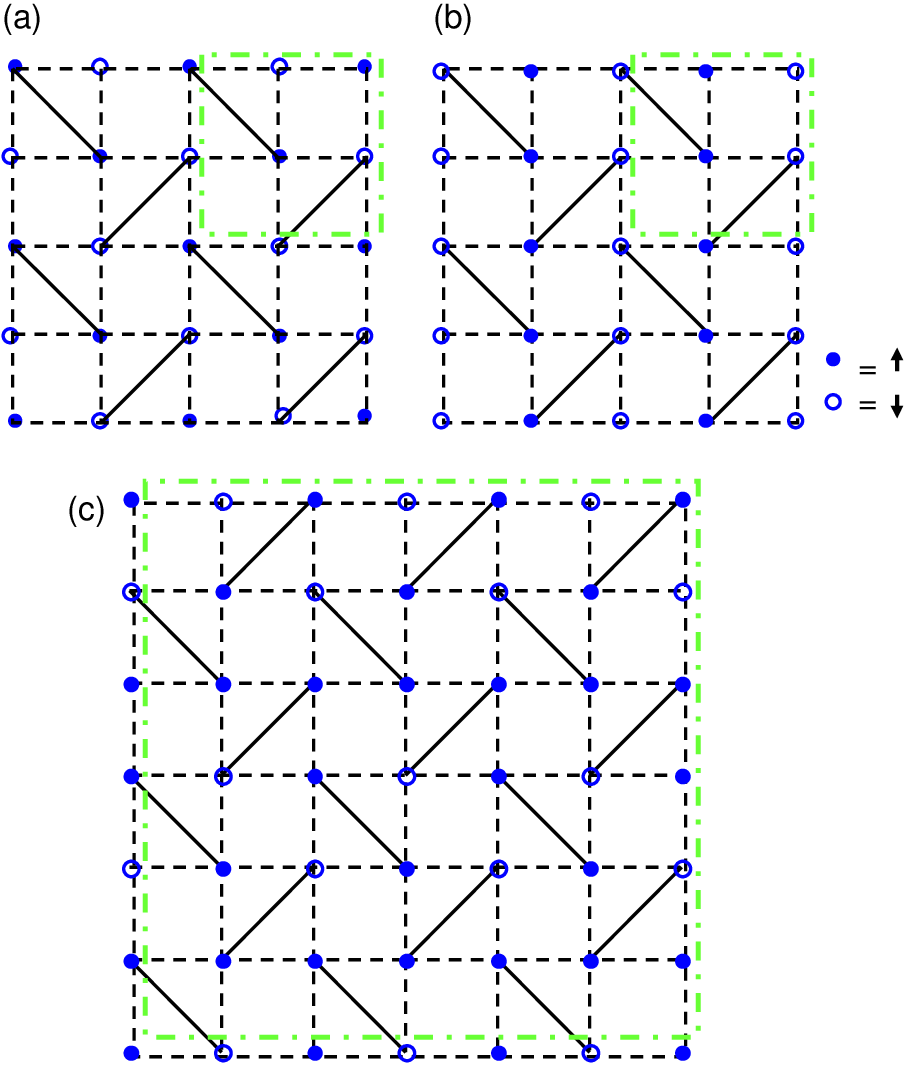}
\caption{(Color online) (a) Spin configuration for the N\'{e}el
state. (b) Spin configuration for the collinear state. (c) Spin
configuration for the magnetization plateau at $m/m_s=1/3$. The
squares with dashed-dotted lines indicate possible choices of unit
cells.}\label{fig:zeroT}
\end{figure}

In the unmagnetized state with $m=0$, we assume that the system is
either in the N\'{e}el state or in the collinear state, depending
on the strength of the frustration $J^\prime$. These states should
be stable when the applied field $h$ is small. When the system is
in the N\'{e}el state [Fig.~\ref{fig:zeroT}(a)], for a unit cell
formed by four plaquettes (bounded by dashed-dotted lines), there
are two sites with spin up and two sites with spin down. The
nearest-neighbor spins are all anti-parallel but the spins
connected by the $J^\prime$-bond are parallel. It is not difficult
to see that the energy per site, including the Zeeman energy (zero
here), is,
\begin{equation}
\epsilon_{m=0}=-\frac{1}{2}+\frac{J^\prime}{8} \; , \label{neel}
\end{equation}
in which $J=1$.

For large frustration, the system is more likely to be in the
collinear state [Fig.~\ref{fig:zeroT}(b)]. Again there are two up
spins and two down spins in a unit cell of four plaquettes. Now
the energy per site becomes
\begin{equation}
{\tilde \epsilon}_{m=0}=-\frac{J^\prime}{8} \; . \label{collinear}
\end{equation}
By comparing the energies in Eqs.~(\ref{neel}) and
(\ref{collinear}), one can see that the energy of the N\'{e}el
state is lower (higher) than the collinear state when $J^\prime
<2$ ($J^\prime >2$).

When the applied field increases, the system can undergo a phase
transition to a $1/3$-plateau state. There are several possible
candidates for such a state. In Fig.~\ref{fig:zeroT}(c), we show
the spin configuration of a state with the lowest possible energy.
With careful analysis, one obtains the following spin energy per
site,
\begin{equation}
\epsilon_{\frac{m}{m_s}=\frac{1}{3}}=-\frac{1}{6}-\frac{J^\prime}{24}-\frac{h}{6}
\;  .
\end{equation}

When the applied field is sufficiently strong, then, irrespective
of the value of $J^\prime$, the system should be fully magnetized.
In such a case, it is relatively easy to determine the spin energy
per site,
\begin{equation}
\epsilon_{\frac{m}{m_s}=1}=\frac{1}{2}+\frac{J^\prime}{8}-\frac{h}{2}
\; . \label{full}
\end{equation}

By comparing $\epsilon_{m=0}$ and $\epsilon_{m/m_s=1/3}$, one can
determine the boundary between the N\'{e}el state and the plateau
state when $J^\prime <2$. The lower critical field $h_{c,1}$ is
found to be
\begin{equation}
h_{c,1}=2-J^\prime \; .
\end{equation}
Similarly, by comparing ${\tilde\epsilon}_{m=0}$ and
$\epsilon_{m/m_s=1/3}$, one has the boundary between the collinear
state and the plateau state when $J^\prime >2$,
\begin{equation}
{\tilde h}_{c,1}=-1+\frac{J^\prime}{2} \; .
\end{equation}
These two straight lines are indicated as the lower phase
boundaries at zero temperature in Fig.~\ref{fig:J'-h_diagram}(b).

On the other hand, the upper critical field $h_{c,2}$ is obtained
by comparing the energies of the plateau state
($\epsilon_{m/m_s=1/3}$) and the fully magnetized state
($\epsilon_{m/m_s=1}$),
\begin{equation}
h_{c,2}=2+\frac{J^\prime}{2} \; .
\end{equation}
Such a straight line is also shown in
Fig.~\ref{fig:J'-h_diagram}(b). The area bounded by these critical
magnetic fields should be the maximum width of the plateau when
the temperature of the system drops to zero. For example, when
$J^\prime =1$, the plateau is bounded by
$(h_{c,1},h_{c,2})=(1,5/2)$ at $T=0$. This agrees nicely with the
extrapolation in Fig.~\ref{fig:T-h_diagram}(b).

\section{Conclusion}

The TRG method is applied to explore the plateau in the
magnetization process for the classical Ising model on the
Shastry-Sutherland lattice.  Systems as large as $2^{10}\times
2^{10}$ sites can be routinely studied with relative ease.
Therefore, the complications from the finite-size effect and its
related geometric frustration can essentially be avoided. We found
a single plateau at $m/m_s=1/3$ that is robust over certain ranges
of temperature and magnetic frustration, consistent with the
result in Ref.~\onlinecite{Meng08} for smaller systems and higher
temperatures. The model under investigation is relevant to the
compound TmB$_4$,~\cite{Siemensmeyer08} which is found to have a
sequence of plateaus down to small fractional
values.~\cite{Gabani08,Siemensmeyer08} We note that the
antiferromagnetic transverse exchanges have not been taken into
account in the current classical model. Therefore, the quantum
effect caused by these couplings may be essential in a full
explanation of the observed plateaus.

\begin{acknowledgments}
M.C.C. thanks the support from the National Science Council of
Taiwan under Contract No. NSC 96-2112-M-003-010-MY3.  MFY
acknowledges the support by the National Science Council of Taiwan
under NSC 96-2112-M-029-004-MY3.
\end{acknowledgments}

\end{document}